\input{aipcheck}

\documentclass[
    ,final            
  ]
  {aipproc}

\layoutstyle{8x11single}

\begin{document}

\title{Optical flashes, reverse shocks and magnetization}

\classification{98.70.Rz}
\keywords      {gamma-ray sources; gamma-ray bursts}

\author{Andreja Gomboc}{
  address={Faculty of Mathematics and Physics, University of Ljubljana, Jadranska 19, SI-1000 Ljubljana, Slovenia}
}
\author{Shiho Kobayashi}{
address={Astrophysics Research Institute, Liverpool John Moores University, Twelve Quays House, Egerton Wharf, Birkenhead, CH41 1LD, UK}
}
\author{Carole G. Mundell}{ 
address={Astrophysics Research Institute, Liverpool John Moores University, Twelve Quays House, Egerton Wharf, Birkenhead, CH41 1LD, UK}
}
\author{Cristiano Guidorzi}{ 
address={Astrophysics Research Institute, Liverpool John Moores University, Twelve Quays House, Egerton Wharf, Birkenhead, CH41 1LD, UK}
 ,altaddress={INAF-Osservatorio Astronomico di Brera, via Bianchi 46, I-23807 Merate (LC), Italy}
 }
\author{Andrea Melandri}{ 
address={Astrophysics Research Institute, Liverpool John Moores University, Twelve Quays House, Egerton Wharf, Birkenhead, CH41 1LD, UK}
}
\author{Iain A. Steele}{ 
address={Astrophysics Research Institute, Liverpool John Moores University, Twelve Quays House, Egerton Wharf, Birkenhead, CH41 1LD, UK}
}
\author{Robert J. Smith}{ 
address={Astrophysics Research Institute, Liverpool John Moores University, Twelve Quays House, Egerton Wharf, Birkenhead, CH41 1LD, UK}
}\author{David Bersier}{address={Astrophysics Research Institute, Liverpool John Moores University, Twelve Quays House, Egerton Wharf, Birkenhead, CH41 1LD, UK}
}
\author{David Carter}{address={Astrophysics Research Institute, Liverpool John Moores University, Twelve Quays House, Egerton Wharf, Birkenhead, CH41 1LD, UK}
}
\author{Michael F. Bode}{address={Astrophysics Research Institute, Liverpool John Moores University, Twelve Quays House, Egerton Wharf, Birkenhead, CH41 1LD, UK}
}

\begin{abstract}
 Despite the pre-Swift expectation that bright optical flashes from reverse shocks would be prevalent in early-time afterglow emission, rapid response observations show this not to be the case. Although very bright at early times, some GRBs such as GRB~061007 and GRB~060418, lack the short-lived optical flash from the reverse shock within minutes after the GRB. In contrast, other optical afterglows, such as those of GRB~990123, GRB~021211, GRB~060111B, GRB~060117, GRB~061126, and recently GRB~080319B, show a steep-to-flat transition within first $10^3$~s typical of a rapidly evolving reverse + forward shock combination. We review the presence and absence of the reverse shock components in optical afterglows and discuss the implications for the standard model and the magnetization of the fireball. We show that the previously predicted optical flashes are likely to occur at lower wavelengths, perhaps as low as radio wavelengths and, by using the case of GRB~061126 we show that the magnetic energy density in the ejecta, expressed as a fraction of the equipartion value, is a key physical parameter.
\end{abstract}

\maketitle


\section{Theory of optical flashes and light curve types}

The theory of GRB afterglows predicts that when the shell of the
relativistically expanding fireball collides with the surrounding
medium, a forward shock is formed which, via synchrotron radiation,
produces the afterglow emission \cite{Sar1995,Mes1997,Sar1998}. In
addition, a reverse shock, propagating backwards through the shell, is
formed. This reverse shock is predicted, under certain conditions, to produce
a bright optical emission - often termed an optical flash
\cite{Sar1999a}.
  
 In studying the evolution of relativistic shells and resultant
afterglows, one can consider two limiting cases: (i) the thick, low
density shell case in which the reverse shock quickly becomes
relativistic and begins to decelerate the shell material; and (ii) the
thin, high density shell case, in which the reverse shock is too weak
to decelerate the shell effectively and the reverse shock does not
become relativistic before the shell crossing time (i.e. it remains
Newtonian or sub-relativistic). The resulting light curves of the
reverse shock in both cases are shown in Figs.~2 and 3 in
\cite{Kob2000}.  
  
\begin{figure}\label{gomboc_fig1}
  \includegraphics[height=.5\textheight]{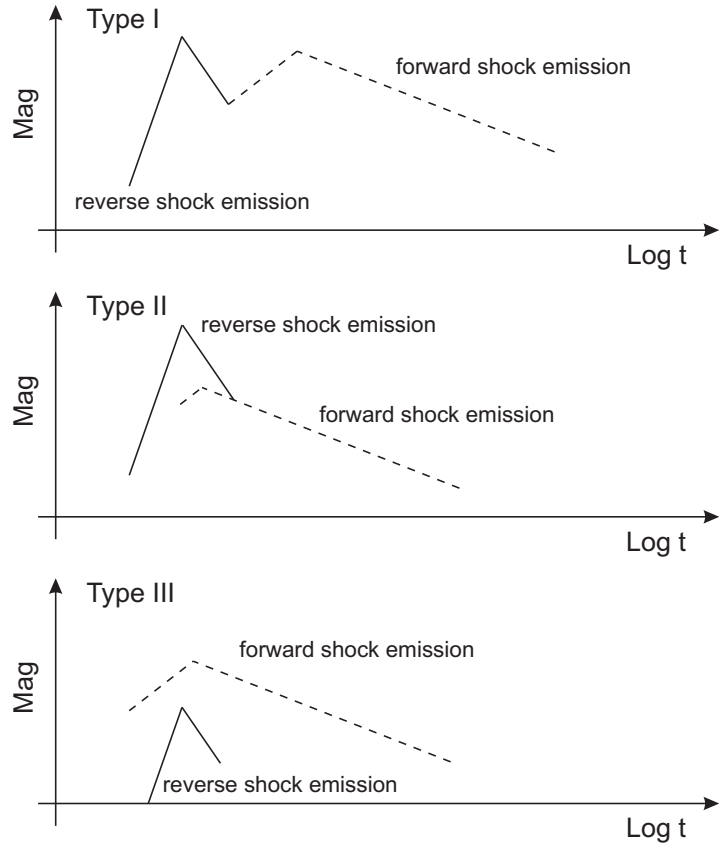}
  \caption{Three types of GRB early optical afterglows, produced as a composition of the reverse and forward shock emission \cite{Zha2003,Jin2007}. Solid line represents reverse shock emission, and dashed line the forward shock emission.}
\end{figure}

 When combining the predicted reverse shock emission with the
afterglow emission from the forward shock, one obtains three types of
light curves \cite{Zha2003,Jin2007} (see Fig.~\ref{gomboc_fig1}): 
  \begin{itemize}
  \item{in Type I, the peak of reverse shock is comparable to or higher
than that of the forward shock. The peaks are well separated in time,
which makes them both clearly visible.}  
  \item{in Type II, the peak of the forward shock is weaker and is
hidden by the bright reverse shock emission. The forward shock
component becomes evident only later, when the reverse shock emission
has faded.} 
  \item{in Type III, the reverse shock emission is
fainter and it is hidden by the forward shock emission at all times.} 
  \end{itemize}
  The theory \cite{Zha2005,Fan2004} predicts that the strength of the
reverse shock emission depends on magnetization (and whether the
typical frequency is close to the optical band). The commonly used
parameter for magnetization is $\sigma$, which is the ratio between
the magnetic and kinetic energy flux. For $\sigma\ll 1$, the jet is
baryonic and magnetic fields are assumed to be produced via local
instabilities at shocks. If the magnetic field originates at the central
engine, and advect outwards with the expanding outflow, $\sigma$ could
be large. The reverse-shock peak flux increases with $\sigma$
initially and starts to decrease when $\sigma \sim$ 1. Therefore, in
order to obtain bright reverse shock emission, $\sigma$ should be
close to unity. 

\section{Observations and the lack of optical flashes?}

\subsection{Detections in the pre-Swift Era}

Before the launch of Swift, there were three optical afterglows
detected, which showed the reverse shock emission signature:  
\begin{itemize}
\item{{\bf GRB~990123} was the most famous one: its bright optical
flash peaked at 8.9~mag \cite{Ake1999} and it was interpreted as due
to the reverse shock, decaying with $\alpha\sim 2$ ($F\propto
t^{-\alpha}$). From about $10^3$~s after the trigger onwards, a more slowly
decaying forward shock component with $\alpha \sim 1.1$ became
dominant \cite{Sar1999b,Mes1999,Nak2005}. According to the above
classification, this was a burst of Type II.} 
\item{{\bf GRB~021004} had an early peak followed by a later
re-brightening at about 0.1~days after the trigger. The features in
the light curve
were modeled with refreshed shocks \cite{Zha2002} and the effects of
density enhancements in the ambient circumburst medium  \cite{Laz2002}.
The light curve was also consistent with a reverse-shock peak followed
by a forward-shock peak \cite{Kob2003} and would therefore belong in
Type I class.} 

\item{{\bf GRB~021211}: no peak was detected in this case, but the
optical light curve underwent a flattening from $\alpha \sim 1.6$ to
$\sim 0.8$ at about 11~min after the trigger
\cite{Fox2003,Pan2003,Wei2003}. It could be interpreted as a Type II
reverse plus forward shock afterglow, but with no reverse shock peak
caught.} 
\end{itemize}

After the detection of the bright optical flash of GRB~990123,
expectations of early time observations were high (however, see also
\cite{Ake2000}). Especially in the Swift era, detections of numerous
bright optical flashes were expected, due to rapid and accurate GRB
localizations by Swift being fed to large ground-based robotic
telescopes capable of immediate follow-up observations.

\begin{figure}\label{gomboc_fig2}
  \includegraphics[height=.5\textheight]{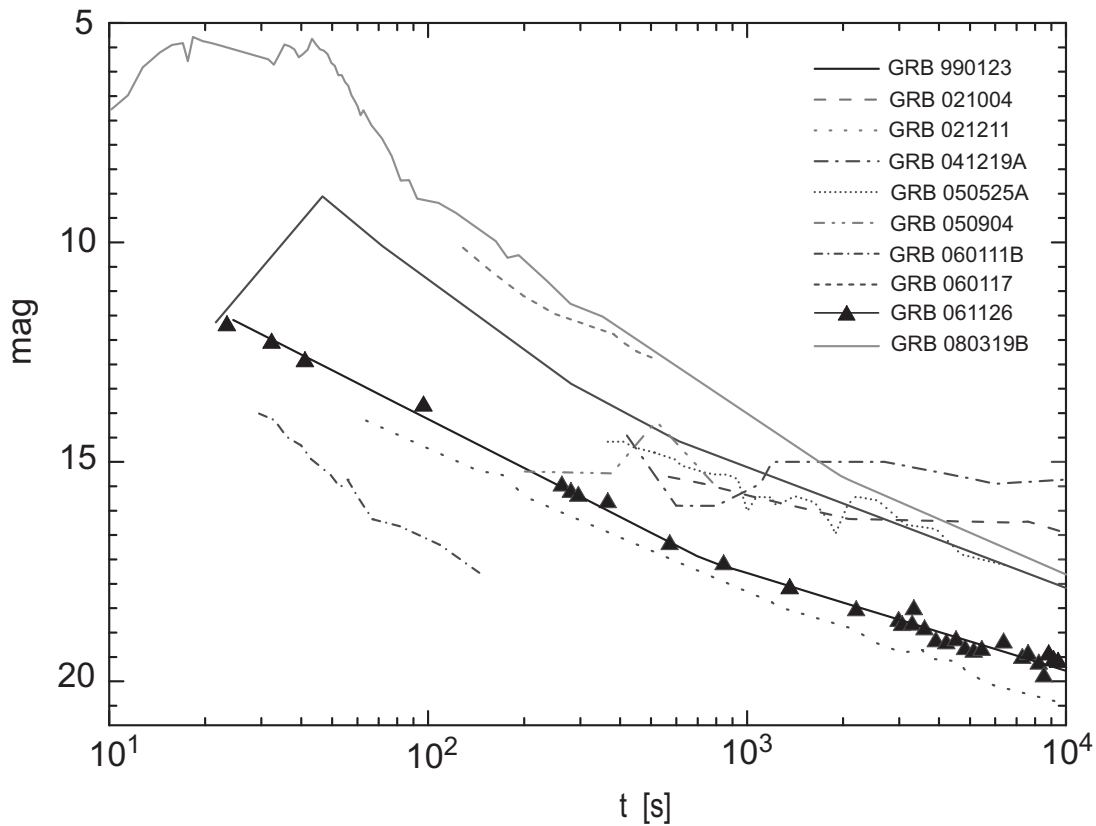}
  \caption{Optical afterglow of GRB~061126. Afterglows of GRB~990123, GRB~021004, GRB~021211, GRB~041219A, GRB~050525A, GRB~050904, GRB~060111B, GRB~060117, and GRB~080319B are shown schematically for comparison.}
\end{figure}

\subsection{Reverse-Shock Detections in the Swift-Era}

Intriguingly, despite high quality ground- and space-based followup
observations, there was a dearth of optical flashes. A study of early
Swift observations with UVOT offered some interpretations and drove
changes in observing strategies \cite{Rom2006}. More recently, we
analysed a sample of 63 GRBs observed between October 2004 and July
2007 by 2-m robotic Liverpool and Faulkes Telescopes\footnote{For
details see \cite{Gom2006} and
\url{http://www.astro.livjm.ac.uk/RoboNet/}.} presented in
\cite{Mel2008}. In this sample, 24 optical afterglows were detected,
but only one - GRB~061126 (see below) - optical emission consistent
with a reverse shock signature. The remaining 39 GRBs had no
detectable optical afterglows despite deep, early time optical
observations in red filters.  The key question therefore remains- why do
bright optical flashes remain elusive?

A full summary of detections of the optical afterglows with a possible
reverse shock component in the Swift Era is as follows:
\begin{itemize} \item{{\bf GRB~041219A} was studied in
\cite{Bla2005,Fan2005}, who showed that its early optical-IR emission
could be understood as a sum of reverse and forward shock emission of
Type I.}  
\item{{\bf GRB~050525A} could also be classified as of Type
I, if its rebrightening at 0.01-0.03~days after the trigger \cite{Klo2005} is
attributed to the forward shock peak \cite{Sha2005,Blu2006}.}  
\item{{\bf GRB~050904} was at $z$=6.3 \cite{Tag2005} one of the most distant GRBs detected. In spite of its distance, it showed a bright optical afterglow, which could be explained with the external forward-reverse shock model or with the "late internal shock model" \cite{Wei2006,Boe2006}.}
\item{{\bf GRB~060111B} had an optical light curve of Type II: a fast decaying phase with $\alpha\sim 1.4$ dominated from 28~s to 80~s after the trigger, and was followed by a shallower decay with $\alpha\sim 1.1$ \cite{Klo2006}.}
\item{{\bf GRB~060117} had a flattening optical light curve, which could be
explained as Type II, summing reverse and forward shock emission
(with $\alpha_{\rm r} \sim 2.5$ and $\alpha_{\rm f} \sim 1.5$) and no
reverse shock {\em peak} detected \cite{Jel2006}.}  
\item{{\bf GRB~061126}
also showed a flattening, Type II, optical light curve, which was well
covered by observations from 21~s to 15~days after the trigger (see
Fig.~\ref{gomboc_fig2}). This case is presented in detail in
\cite{Per2008} and \cite{Gom2008}. Although observations started very
early, they did not catch the rise of the afterglow. The light curve
at early time however shows the steeply decaying reverse-shock component to
$\sim$13~min after the trigger. After this time, the light curve is dominated by the forward
shock emission. Analysis presented in \cite{Gom2008} shows that
the decay indices of the reverse and forward shocks of $\alpha_{\rm r}=1.69 \pm0.09$ and
$\alpha_{\rm f}=0.78\pm0.04$ respectively are in agreement with those
expected from the standard model for $p=2$, which follows from the
observed $\beta_{\rm X} \sim 1$. However, the X-ray afterglow
decays faster than the optical (with $\alpha_{\rm X} \sim 1.3$) and
is not in agreement with the standard model. This could indicate
a different origin of the X-ray afterglow, which would perhaps also
explain two other intriguing properties of this burst: (i) its optical
"darkness" at early time: the ratio of optical to X-ray flux places this
burst in the "dark burst" region in the
Jakobbson plot \cite{Jak2004}, despite its afterglow being bright,
$R \sim 12$~mag, at the beginning of observations, and (ii) a possible
chromatic break: a break seen in optical light curve a few days
after the trigger that absent from the X-ray light curve, which showed
no break.}  
\item{{\bf GRB~080319B} was a recent case of an
optical flash, the well known naked-eye burst \cite{Rac2008}. The
observational coverage of its optical emission was unprecedented. It showed Type II
characteristics and two steep-to-flat transitions from $\alpha\sim 6.5$
to $\alpha\sim 2.5$ at about 100~s after the trigger, to
$\alpha\sim 1.25$ at about $10^3$~s after the trigger. As this is a
well-studied burst, we refer the reader to e.g. \cite{Rac2008} for
more detail.}  \end{itemize}

\subsection{GRBs lacking Reverse Shock Emission in the Swift-Era}

In addition to GRBs with bright afterglows consistent with reverse
shock emission at early time, GRBs with bright optical counterparts
have also been detected that are not consistent with reverse shock
emission.

One explanation for the absence of bright reverse-shock emission is
suppression due to a high degree of magnetization of the fireball. In
the case of GRB~060418, which showed Type III optical afterglow, there
is another option to explain the absence of a distinctive reverse
shock component. Even in the standard model, if the typical frequency
of the forward shock is already below the optical band at the shock
crossing time, the two shock emissions peak at the same time, and they
contribute equally to the observed optical light. The superposition
produces a single peak.  Using the RINGO polarimeter \cite{Ste2006}
mounted on the Liverpool Telescope, an automatic measurement of the early-time
polarization of the GRB~060418 afterglow was taken at its peak, i.e. at the
onset of the afterglow \cite{Mun2007a}. A low upper limit of 8\% shows
that no large-scale ordered magnetic field is actually needed or that
very strong magnetic field $\sigma \gg 1$ suppresses the reverse shock
emission.

A low typical frequency of the reverse shock, which is lower by a
factor of $\Gamma^2$ than that of forward shock, also explains the
lack of optical flash in GRB~061007 \cite{Mun2007b}. The bright
optical afterglow (R$>$10.3 mag) was observed from 137~s to $10^5$~s
after the trigger and showed only a simple decay with $\alpha \sim
1.7$. This afterglow was explained with forward shock emission,
while the typical frequency of the reverse shock emission lies in
radio wavebands already at the beginning of observations at 137~s
after the trigger.

Other possible explanations for the paucity of optical flashes that
have been suggested are: dust extinction
\cite{Sod2002,Sta2007,Sch2007,Li2008}, and the inverse Compton
process, in which most of the energy is emitted through IC process and
synchrotron component is suppressed \cite{Kob2007}. Most commonly,
though, strong suppression of the reverse shock in the Poynting-flux
dominated outflow is considered \cite{Zha2005,Fan2004,Miz2008}. Some
new theoretical results seem to indicate that even a mild
magnetization can suppress a reverse shock
\cite{Mim2008,Fan2008,Gia2008}.   

\newpage
\section{Observations and Magnetization}

We may use those GRBs with reverse shock signature to study
the magnetization of the fireball.  We define parameter
$R_B=\epsilon_{B,r}/\epsilon_{B,f}$ as the ratio of magnetic
equipartition parameters in the reverse shock and forward shock (Note
that the definition of the magnetization is different from that in
\cite{Zha2003}.). Results for some of the above mentioned GRBs are the
following: \begin{itemize} \item{GRB~990123: $R_B \sim $ 200
\cite{Zha2003},} \item{GRB~021004: $R_B \sim $ 1 \cite{Zha2003},}
\item{GRB~021211: $R_B\geq$ 1 \cite{Zha2003,Kum2003},}
\item{GRB~041219A: $R_B \sim $ 10 \cite{Fan2005},} \item{GRB~061126:
$R_B \sim$ 50 \cite{Gom2008}.}  \end{itemize}

These results imply that magnetic energy density in the reverse shock
is  equal to or larger than that in the forward shock. Nevertheless, since
$\epsilon_{B,f} \sim  10^{-4} - 10^{-2}$ is usually inferred from
afterglow modeling \cite{Pan2002}, it follows that $\epsilon_{B,r} <
1$. Therefore, in all the above cases we are dealing with mildly
magnetized outflow and not a Poynting flux dominated outflow. 

\subsection{Polarization}
Theoretical models predict that mildy magnetized outflows produce
strong reverse shock emission and that this emission should be
polarized \cite{Lyu2003,Gra2003,Ros2004,Fan2004,Cov2007}.  
Therefore, polarization measurements at very early times are of key
importance to distinguish between the usual baryon-rich fireball model
and the Poynting flux-dominated outflow model.  

The predicted very early afterglow (possibly even an optical flash)
should be strong enough not only to be detected by current robotic
telescopes, but also to make polarimetry
measurements possible. This was the motivation behind the RINGO polarimeter on
the Liverpool Telescope: the Liverpool Telescope is able to commence
observations in $\sim$ 100~s after the trigger and is capable of good
polarimetric accuracy for bright targets  (for details see
\cite{Ste2006}). The feasibility of such early polarimetry was
demonstrated in the case of GRB~060418 \cite{Mun2007a}. This burst was
of Type III, with hidden reverse shock component. It would be of even
greater interest to measure the early polarization of an afterglow
with the prominent reverse shock contribution or even during the
reverse shock peak (Type I or II). This would provide further
additional constraints on the presence of magnetized fireballs. 

\section{Conclusions} The absence of a large number of optical flashes
might be explained with the standard reverse shock model via weak,
non-relativistic reverse shocks and a typical frequency well below the
optical band \cite{Kob2000}. Some optical flashes may be suppressed by
a strong magnetic field, although new theoretical results suggest that
even mild magnetization may suppress a reverse shock. Catching more
GRBs with reverse shock components is important for distinguishing
between these possibilities, understanding reverse shocks and the role
of magnetization. Magnetization is a fundamental and yet unsolved issue of
GRB physics. Early polarimetry may be of the key importance to help
solve this issue, since it may prove to be a truly needed independent
probe of the physical conditions of the GRB afterglow.


\begin{theacknowledgments}
 AG thanks Slovenian Research Agency and Slovenian Ministry for Higher Education, Science, and Technology for financial support.
 C.G. acknowledges support from ASI grant I/011/07/0. 
 CGM acknowledges financial support from the Royal Society and Research
 Councils U. K. 
 {\it RoboNet-1.0} was supported by PPARC and STFC. {\em Swift} mission is funded in the UK by STFC, in Italy by ASI, and in the USA by NASA.  
 The Faulkes Telescopes are operated by the Las Cumbres Observatory.
 The Liverpool Telescope is owned and operated by Liverpool John Moores University.
\end{theacknowledgments}



\end{document}